\documentclass[a4paper,amsmath,amssymb,amsfonts]{jpconf}
\usepackage{graphicx}

\begin{document}
\title{Wormhole geometries in modified gravity}

\author{Francisco S.~N.~Lobo}

\address{Centro de Astronomia
e Astrof\'{\i}sica da Universidade de Lisboa, Campo Grande, Ed. C8
1749-016 Lisboa, Portugal}

\ead{flobo@cii.fc.ul.pt}

\begin{abstract}
 A fundamental ingredient in wormhole physics is the presence of exotic matter, which involves the violation of the null energy condition. Although a plethora of wormhole solutions have been explored in the literature, it is useful to find geometries that minimize the usage of exotic matter. In the context of modified gravity, it has also been shown that the normal matter can be imposed to satisfy the null energy condition, and it is the higher order curvature terms, interpreted as a gravitational fluid, that sustain these non-standard wormhole geometries, fundamentally different from their counterparts in general relativity. In this paper, we review recent work in wormhole physics in the context of modified theories of gravity.
\end{abstract}

\section{Introduction}

Traversable wormholes are hypothetical tunnels in spacetime given by the following line 
element
\begin{equation}
ds^{2}=-e^{2\Phi(r)}dt^{2}+\left[1-\frac{b(r)}{r}\right]^{-1}\,dr^2+r^{2}(d\theta ^{2}+\sin
^{2}\theta d\phi ^{2}) \,,
  \label{WHmetric}
\end{equation}
where $\Phi(r)$ and $b(r)$ are arbitrary functions of the radial coordinate, $r$, denoted as 
the redshift function, and the shape function, respectively \cite{Morris:1988cz}. The radial 
coordinate $r$ decreases from infinity to a minimum value $r_0$, representing the location of 
the wormhole throat, where $b(r_0)=r_0$, and then increases from $r_0$ back to infinity. For 
the wormhole to be traversable, one must demand that there are no horizons present, which are 
identified as the surfaces with $e^{2\Phi}\rightarrow0$, so that $\Phi(r)$ must be finite 
everywhere. Note that in any region where the $t$ coordinate is timelike requires that $r > 
b(r)$. 

Now, a fundamental property of wormholes is the flaring out of the throat, which 
is  translated by the condition $(b-b^{\prime}r)/b^{2}>0$ \cite{Morris:1988cz}. At 
the throat $b(r_{0})=r=r_{0}$, the condition $b'(r_{0})<1$ is imposed in order to have 
wormhole solutions. It is precisely these restrictions that impose the violation of the null 
energy condition (NEC) in classical general relativity. At this stage, it is important to 
emphasize that the NEC violation involves a stress-energy tensor such that $T_{\mu\nu}k^\mu 
k^\nu <0$, where $k^\mu$ is any null vector \cite{Morris:1988cz}.

However, recently it has been shown in the context of modified theories of gravity that the 
matter threading the wormhole may satisfy the energy conditions, and it is precisely the 
effective stress-energy tensor involving higher order derivatives that is responsible for the 
violation of the NEC. This approach has been explored in $f(R)$ gravity \cite{modgravity1}, 
curvature-matter couplings \cite{modgravity2a,modgravity2b}, conformal Weyl gravity 
\cite{modgravity3} and in braneworlds \cite{modgravity4}, amongst other contexts. In this 
paper, we review some of these scenarios.

\section{Wormholes in modified theories of gravity}

\subsection{$f(R)$ modified theories of gravity}

In first place, we consider $f(R)$ modified theories of gravity, which has recently been 
extensively explored as a possible cause of the late-time cosmic acceleration (see 
\cite{Lobo:2008sg} for a review). The action is given by
\begin{equation}
S=\frac{1}{2\kappa^2}\int d^4x\sqrt{-g}\;f(R)+S_M(g^{\mu\nu},\psi)
\,,
\end{equation}
where $\kappa^2 =8\pi G$, and for notational simplicity we consider $\kappa^2=1$. 
$S_M(g^{\mu\nu},\psi)$ is the matter action, defined as $S_M=\int d^4x\sqrt{-g}\;{\cal 
L}_m(g_{\mu\nu},\psi)$, where ${\cal L}_m$ is the matter Lagrangian density, in which matter 
is minimally coupled to the metric $g_{\mu\nu}$ and $\psi$ collectively denotes the matter
fields.

Varying the action with respect to the metric, one deduces the gravitational field 
equation written in the following form
\begin{equation}
G_{\mu\nu}\equiv R_{\mu\nu}-\frac{1}{2}R\,g_{\mu\nu}= T^{{\rm
eff}}_{\mu\nu} \,,
    \label{gravfield}
\end{equation}
where the effective stress-energy tensor, $T^{{\rm eff}}_{\mu\nu}$, is given by
\begin{eqnarray}
T^{{\rm eff}}_{\mu\nu}= \frac{1}{F}\left[T^{(m)}_{\mu\nu}+\nabla_\mu \nabla_\nu F
-\frac{1}{4}g_{\mu\nu}\left(RF+\nabla^\alpha \nabla_\alpha F+T\right) \right]    \,,
    \label{gravfluid}
\end{eqnarray}
where $F=df/dR$.

The flaring out condition for wormhole geometries imposes $T^{{\rm eff}}_{\mu\nu} \, k^\mu 
k^\nu< 0$, which yields the following generic condition in $f(R)$ gravity
\begin{eqnarray}
\frac{1}{F}\left(T^{(m)}_{\mu\nu}\, k^\mu k^\nu +\, k^\mu k^\nu \nabla_\mu \nabla_\nu F 
\right) < 0   \,.
    \label{NECgravfluid}
\end{eqnarray}
This condition has been extensively explored in \cite{modgravity1}, and specific wormhole 
solutions have been found. In principle, one may impose the condition $T^{(m)}_{\mu\nu} k^\mu 
k^\nu\ge 0$ for the normal matter threading the wormhole. This is extremely useful, as 
applying local Lorentz transformations it is possible to show that the above condition implies 
that the energy density is positive in all local frames of reference. Note that in general 
relativity, i.e., $f(R)=R$, we regain the condition for the matter stress-energy tensor 
violation of the NEC, i.e., $T^{(m)}_{\mu\nu} \, k^\mu k^\nu< 0$.

Thus, in the context of $f(R)$ modified theories of gravity it is the higher order curvature 
terms, interpreted as a gravitational fluid, that sustain these non-standard wormhole 
geometries, fundamentally different from their counterparts in general relativity. 

\subsection{Curvature-matter coupling}

An interesting generalization of $f(R)$ gravity is a model described by a curvature-matter 
coupling considered in \cite{Koivisto,Bertolami}. The action is given by
\begin{equation}
S=\int \left\{\frac{1}{2}f_1(R)+\left[1+\lambda f_2(R)\right]{\cal
L}_{m}\right\} \sqrt{-g}\;d^{4}x~\,,
\end{equation}
where $f_i(R)$ (with $i=1,2$) are arbitrary functions of the Ricci scalar $R$ and ${\cal 
L}_{m}$ is the Lagrangian density corresponding to matter. The coupling constant $\lambda$ 
characterizes the strength of the interaction between $f_2(R)$ and the matter Lagrangian.

The effective gravitational field equation is given by Eq. (\ref{gravfield}), with the 
effective stress-energy tensor provided by
\begin{eqnarray}
T^{{\rm eff}}_{\mu\nu}= \left[1+\lambda f_2(R)\right]T_{\mu \nu }^{(m)}
-2\lambda \left[ F_2(R){\cal L}_m R_{\mu\nu}
   -(\nabla_\mu
\nabla_\nu-g_{\mu\nu}\nabla^\alpha \nabla_\alpha){\cal L}_m F_2(R)\right]\,,
    \label{efffield2}
\end{eqnarray}
where for simplicity we have considered $f_1(R)=R$.

Imposing $T^{(m)}_{\mu\nu} k^\mu k^\nu\ge 0$ for the normal matter threading the wormhole, the 
NEC violation condition $T^{{\rm eff}}_{\mu\nu} \, k^\mu k^\nu< 0$, and considering that 
$1+\lambda f_2(R)>0$, imposes the restriction
\begin{eqnarray}
0\leq T_{\mu \nu }^{(m)}k^{\mu}k^{\nu}< \frac{2\lambda}{1+\lambda f_2(R)}
\big[ F_2(R){\cal L}_m \, R_{\mu\nu}\,k^{\mu}k^{\nu}
   -k^{\mu}k^{\nu}\,\nabla_\mu
\nabla_\nu {\cal L}_m F_2(R)\big]\,.
    \label{effNEC1}
\end{eqnarray}

In an analogous manner to $f(R)$ modified theories of gravity, it is the higher order 
curvature terms in the curvature-matter coupling that sustain these non-standard wormhole 
geometries. These conditions have been extensively analysed and specific solutions have been 
found in \cite{modgravity2a,modgravity2b}.

\section{General class of braneworld wormholes}

Braneworld cosmology is an interesting scenario based on the idea that our Universe is a 
$3-$brane embedded in a five-dimensional bulk. Considering that the bulk is vacuous, the 
induced field equations on the brane may formally be rewritten by Eq. (\ref{gravfield}), where 
the effective stress-energy tensor is given by
\begin{equation}\label{effstress}
T_{\mu\nu}^{\rm eff} =T_{\mu\nu}-\frac{1}{\kappa^2}\,{\cal E}_{\mu\nu}
+ \frac{6}{\lambda}\,S_{\mu\nu} \,,
\end{equation}
with $\kappa^2=\lambda k_5^2/6$; $\kappa^2$ and $k_5^2$ are the gravitational coupling  
constants, on the brane and in the bulk, respectively; $\lambda$ is the tension on the brane 
and $T_{\mu\nu}$ is the stress energy tensor confined on the brane. The first correction 
term relative to Einstein's general relativity is the inclusion of a quadratic term 
$S_{\mu\nu}$ in the stress-energy tensor, arising from the extrinsic curvature terms
in the projected Einstein tensor, and is given by
\begin{equation}
S_{\mu\nu}=\frac{1}{12}T T_{\mu\nu}-\frac{1}{4}
T_{\mu\alpha}T^{\alpha}{}_{\nu}+
\frac{1}{8}\,g_{\mu\nu}\left[T_{\alpha\beta}T^{\alpha\beta}-\frac{1}{3}T^2
\right] \,,
    \label{inducedEFE}
\end{equation}
with $T=T^{\mu}{}_{\mu}$.
The second correction term, ${\cal E}_{\mu\nu}$, is the projection
of the 5-dimensional Weyl tensor onto the
brane, and encompasses nonlocal bulk effects. The only general known
property of this nonlocal term is that it is traceless, i.e.,
${\cal E}^{\mu}{}_{\mu}=0$.

We have considered, for simplicity, that the
cosmological constant on the brane is zero. Note that the
quadratic term, i.e., $S_{\mu\nu}\sim (T_{\mu\nu})^2$, is the high
energy correction term. From the following approximations
$|S_{\mu\nu}/\lambda|/|T_{\mu\nu}|\sim |T_{\mu\nu}|/\lambda\sim
\rho/\lambda$, one readily verifies that $S_{\mu\nu}$ is dominant
for $\rho\gg \lambda$, and negligible in the regime $\rho\ll
\lambda$, where $\lambda > (1 {\rm Tev})^4$ \cite{branereview}.

Now, the condition $T^{{\rm eff}}_{\mu\nu} \, k^\mu k^\nu< 0$, yields the following generic condition in braneworld gravity
\begin{equation}\label{effstressNEC}
0 \leq T_{\mu\nu}k^\mu k^\nu < \frac{1}{\kappa^2}\,{\cal E}_{\mu\nu}\, k^\mu k^\nu
- \frac{6}{\lambda}\,S_{\mu\nu}\, k^\mu k^\nu \,,
\end{equation}
Thus, imposing that the stress energy tensor confined on the brane, and threading the 
wormhole, satisfies the NEC, we verify that in addition to the local high-energy bulk effects, 
nonlocal corrections from the Weyl curvature in the bulk may induce a NEC violating signature 
on the brane. Thus, braneworld gravity seems to provide a natural scenario for the existence 
of traversable wormholes.

\section{Conclusion}

In this work, we have explored the possibility that wormholes be supported by modified 
theories of gravity. We imposed that the matter threading the wormhole satisfies the energy 
conditions, and it is the higher order curvature derivative terms (in the context of $f(R)$ 
gravity and its curvature-matter coupling generalization) and the local high-energy bulk 
effects and nonlocal corrections from the Weyl curvature in the bulk (in braneworld gravity), 
that support these nonstandard wormhole geometries, fundamentally different from their 
counterparts in general relativity.

\subsection*{Acknowledgments}
The author acknowledges financial support of the Funda\c{c}\~{a}o para a Ci\^{e}ncia e 
Tecnologia through Grants PTDC/\-FIS/\-102742/2008 and CERN/FP/116398/2010.

\end{document}